\documentclass[twocolumn,prc,aps,superscriptaddress,showpacs,floatfix]{revtex4}

\usepackage{graphicx}
\usepackage{comment}

\def\ge{G_E}
\def\gm{G_M}
\def\mugegm{\mu_p G_E/G_M}
\def\gegm{G_E/G_M}
\def\deg{^\circ}
\def\etal{\textit{et al.}}
\def\gtorder{\mathrel{\raise.3ex\hbox{$>$}\mkern-14mu
 \lower0.6ex\hbox{$\sim$}}}
\def\ltorder{\mathrel{\raise.3ex\hbox{$<$}\mkern-14mu
 \lower0.6ex\hbox{$\sim$}}}

\begin{document}
\title{Low $Q^2$ measurements of the proton form factor ratio $\mugegm$}

\author{G.~Ron}
\affiliation{The Weizmann Institute of Science, Rehovot 76100, Israel} 
\affiliation{Lawrence Berkeley National Lab, Berkeley, CA 94720, USA}
\affiliation{Racah Institute of Physics, Hebrew University of Jerusalem, Jerusalem, Israel 91904}
\author{X.~Zhan}
\affiliation{Massachusetts Institute of Technology, Cambridge, Massachusetts 02139, USA}
\author{J.~Glister}
\affiliation{Saint Mary's University, Halifax, Nova Scotia B3H 3C3, Canada}
\affiliation{Dalhousie University, Halifax, Nova Scotia B3H 3J5, Canada}
\author{B.~Lee}
\affiliation{Seoul National University, Seoul 151-747, Korea}
\author{K.~Allada}
\affiliation{University of Kentucky, Lexington, Kentucky 40506, USA}
\author{W.~Armstrong}
\affiliation{Temple University, Philadelphia, Pennsylvania 19122, USA}
\author{J.~Arrington}
\affiliation{Argonne National Laboratory, Argonne, Illinois, 60439, USA}
\author{A.~Beck}
\affiliation{Massachusetts Institute of Technology, Cambridge, Massachusetts 02139, USA}
\affiliation{Thomas Jefferson National Accelerator Facility, Newport News, Virginia 23606, USA}
\author{F.~Benmokhtar}
\affiliation{University of Maryland, Baltimore, Maryland, USA}
\author{B.L.~Berman}
\affiliation{George Washington University, Washington D.C. 20052, USA}
\author{W.~Boeglin}
\affiliation{Florida International University, Miami, Florida 33199, USA}
\author{E.~Brash}
\affiliation{Christopher Newport University, Newport News, Virginia, 2360X, USA}
\author{A.~Camsonne}
\affiliation{Thomas Jefferson National Accelerator Facility, Newport News, Virginia 23606, USA}
\author{J.~Calarco}
\affiliation{University of New Hampshire, Durham, New Hampshire 03824, USA}
\author{J.~P.~Chen}
\affiliation{Thomas Jefferson National Accelerator Facility, Newport News, Virginia 23606, USA}
\author{Seonho~Choi}
\affiliation{Seoul National University, Seoul 151-747, Korea}
\author{E.~Chudakov}
\affiliation{Thomas Jefferson National Accelerator Facility, Newport News, Virginia 23606, USA}
\author{L.~Coman}
\affiliation{}
\author{B.~Craver}
\affiliation{University of Virginia, Charlottesville, Virginia 22904, USA}
\author{F.~Cusanno}
\affiliation{INFN, Sezione Sanit\'{a} and Istituto Superiore di Sanit\'{a}, Laboratorio di Fisica, I-00161 Rome, Italy}
\author{J.~Dumas}
\affiliation{Rutgers, The State University of New Jersey, Piscataway, New Jersey 08855, USA}
\author{C.~Dutta}
\affiliation{University of Kentucky, Lexington, Kentucky 40506, USA}
\author{R.~Feuerbach}
\affiliation{Thomas Jefferson National Accelerator Facility, Newport News, Virginia 23606, USA}
\author{A.~Freyberger}
\affiliation{Thomas Jefferson National Accelerator Facility, Newport News, Virginia 23606, USA}
\author{S.~Frullani}
\affiliation{INFN, Sezione Sanit\'{a} and Istituto Superiore di Sanit\'{a}, Laboratorio di Fisica, I-00161 Rome, Italy}
\author{F.~Garibaldi}
\affiliation{INFN, Sezione Sanit\'{a} and Istituto Superiore di Sanit\'{a}, Laboratorio di Fisica, I-00161 Rome, Italy}
\author{R.~Gilman}
\affiliation{Rutgers, The State University of New Jersey, Piscataway, New Jersey 08855, USA}
\affiliation{Thomas Jefferson National Accelerator Facility, Newport News, Virginia 23606, USA}
\author{O.~Hansen}
\affiliation{Thomas Jefferson National Accelerator Facility, Newport News, Virginia 23606, USA}
\author{D.~W.~Higinbotham}
\affiliation{Thomas Jefferson National Accelerator Facility, Newport News, Virginia 23606, USA}
\author{T.~Holmstrom}
\affiliation{College of William and Mary, Williamsburg, Virginia 23187, USA}               
\author{C.E.~Hyde}
\affiliation{Old Dominion University, Norfolk, Virginia 23508, USA}
\author{H.~Ibrahim}
\affiliation{Old Dominion University, Norfolk, Virginia 23508, USA}
\author{Y. Ilieva}
\affiliation{George Washington University, Washington D.C. 20052, USA}
\author{C.~W.~de~Jager}
\affiliation{Thomas Jefferson National Accelerator Facility, Newport News, Virginia 23606, USA}
\author{X.~Jiang}
\affiliation{Rutgers, The State University of New Jersey, Piscataway, New Jersey 08855, USA}
\author{M.~Jones}
\affiliation{Thomas Jefferson National Accelerator Facility, Newport News, Virginia 23606, USA}
\author{A.~Kelleher}
\affiliation{College of William and Mary, Williamsburg, Virginia 23187, USA}               
\author{E.~Khrosinkova}
\affiliation{Kent State University, Kent, Ohio 44242, USA}
\author{E.~Kuchina}
\affiliation{Rutgers, The State University of New Jersey, Piscataway, New Jersey 08855, USA}
\author{G.~Kumbartzki} 
\affiliation{Rutgers, The State University of New Jersey, Piscataway, New Jersey 08855, USA}
\author{J.~J.~LeRose}
\affiliation{Thomas Jefferson National Accelerator Facility, Newport News, Virginia 23606, USA}
\author{R.~Lindgren}
\affiliation{University of Virginia, Charlottesville, Virginia 22904, USA}
\author{P.~Markowitz}
\affiliation{Florida International University, Miami, Florida 33199, USA}
\author{S.~May-Tal~Beck}
\affiliation{Massachusetts Institute of Technology, Cambridge, Massachusetts 02139, USA}
\affiliation{Thomas Jefferson National Accelerator Facility, Newport News, Virginia 23606, USA}
\author{E.~McCullough}
\affiliation{Saint Mary's University, Halifax, Nova Scotia B3H 3C3, Canada}
\author{M.~Meziane}
\affiliation{College of William and Mary, Williamsburg, Virginia 23187, USA}               
\author{Z.-E.~Meziani}
\affiliation{Temple University, Philadelphia, Pennsylvania 19122, USA}
\author{R.~Michaels}
\affiliation{Thomas Jefferson National Accelerator Facility, Newport News, Virginia 23606, USA}
\author{B.~Moffit}
\affiliation{College of William and Mary, Williamsburg, Virginia 23187, USA}               
\author{B.E.~Norum}
\affiliation{University of Virginia, Charlottesville, Virginia 22904, USA}
\author{Y.~Oh}
\affiliation{Seoul National University, Seoul 151-747, Korea}
\author{M.~Olson}
\affiliation{Saint Norbert College, Greenbay, Wisconsin 54115, USA}
\author{M.~Paolone}
\affiliation{University of South Carolina, Columbia, South Carolina 29208, USA}
\author{K.~Paschke}
\affiliation{University of Virginia, Charlottesville, Virginia 22904, USA}
\author{C.~F.~Perdrisat}
\affiliation{College of William and Mary, Williamsburg, Virginia 23187, USA}               
\author{E.~Piasetzky}
\affiliation{Tel Aviv University, Tel Aviv 69978, Israel} 
\author{M.~Potokar} 
\affiliation{Institute ``Jo\v{z}ef Stefan'', 1000 Ljubljana, Slovenia}
\author{R.~Pomatsalyuk}
\affiliation{Kharkov Institue, Kharkov 310108, Ukraine}
\affiliation{Thomas Jefferson National Accelerator Facility, Newport News, Virginia 23606, USA}
\author{I.~Pomerantz}
\affiliation{Tel Aviv University, Tel Aviv 69978, Israel} 
\author{A.~J.~R.~Puckett}
\affiliation{Massachusetts Institute of Technology, Cambridge, Massachusetts 02139, USA}
\author{V.~Punjabi}
\affiliation{Norfolk State University, Norfolk, Virginia 23504, USA}
\author{X.~Qian}
\affiliation{Duke University, Durham, NC 27708, USA}
\author{Y.~Qiang}
\affiliation{Massachusetts Institute of Technology, Cambridge, Massachusetts 02139, USA}
\author{R.~Ransome}
\affiliation{Rutgers, The State University of New Jersey, Piscataway, New Jersey 08855, USA}
\author{M.~Reyhan}
\affiliation{Rutgers, The State University of New Jersey, Piscataway, New Jersey 08855, USA}
\author{J.~Roche}
\affiliation{Ohio University, Athens, Ohio 45701, USA}
\author{Y.~Rousseau}
\affiliation{Rutgers, The State University of New Jersey, Piscataway, New Jersey 08855, USA}
\author{A.~Saha} 
\affiliation{Thomas Jefferson National Accelerator Facility, Newport News, Virginia 23606, USA}
\author{A.J.~Sarty}
\affiliation{Saint Mary's University, Halifax, Nova Scotia B3H 3C3, Canada}
\author{B.~Sawatzky}
\affiliation{University of Virginia, Charlottesville, Virginia 22904, USA}
\affiliation{Temple University, Philadelphia, Pennsylvania 19122, USA}
\author{E.~Schulte}
\affiliation{Rutgers, The State University of New Jersey, Piscataway, New Jersey 08855, USA}
\author{M.~Shabestari}
\affiliation{University of Virginia, Charlottesville, Virginia 22904, USA}
\author{A.~Shahinyan}
\affiliation{Yerevan Physics Institute, Yerevan 375036, Armenia}
\author{R.~Shneor}
\affiliation{Tel Aviv  University, Tel Aviv 69978, Israel} 
\author{S.~\v{S}irca}
\affiliation{Dept. of Physics, University of Ljubljana, 1000 Ljubljana, Slovenia}
\affiliation{Institute ``Jo\v{z}ef Stefan'', 1000 Ljubljana, Slovenia}
\author{K.~Slifer}
\affiliation{University of Virginia, Charlottesville, Virginia 22904, USA}
\author{P.~Solvignon}
\affiliation{Argonne National Laboratory, Argonne, Illinois, 60439, USA}
\author{J.~Song}
\affiliation{Seoul National University, Seoul 151-747, Korea}
\author{R.~Sparks}
\affiliation{Thomas Jefferson National Accelerator Facility, Newport News, Virginia 23606, USA}
\author{R.~Subedi}
\affiliation{Kent State University, Kent, Ohio 44242, USA}
\author{S.~Strauch}
\affiliation{University of South Carolina, Columbia, South Carolina 29208, USA}
\author{G.~M.~Urciuoli}
\affiliation{INFN, Sezione di Roma, I-00185 Rome, Italy}
\author{K.~Wang}
\affiliation{University of Virginia, Charlottesville, Virginia 22904, USA}
\author{B.~Wojtsekhowski}
\affiliation{Thomas Jefferson National Accelerator Facility, Newport News, Virginia 23606, USA}
\author{X.~Yan}
\affiliation{Seoul National University, Seoul 151-747, Korea}
\author{H.~Yao}
\affiliation{Temple University, Philadelphia, Pennsylvania 19122, USA}
\author{X.~Zhu}
\affiliation{Duke University, Durham, North Carolina 27708, USA}
\collaboration{The Jefferson Lab Hall A Collaboration}
\noaffiliation
\date{\today}


\bibliographystyle{apsrev}


\begin{abstract}

We present an updated extraction of the proton electromagnetic form factor
ratio, $\mugegm$, at low $Q^2$.  The form factors are sensitive to the spatial
distribution of the proton, and precise measurements can be used to constrain
models of the proton. An improved selection of the elastic events and reduced
background contributions yielded a small systematic reduction in the ratio
$\mugegm$ compared to the original analysis.

\end{abstract}

\maketitle


\section{Introduction}\label{sec:introduction}

We present a detailed reanalysis of polarization transfer measurements of the
proton form factor ratio $\mugegm$ initially presented in Ref.~\cite{ron07},
with improved selection of elastic events and significantly reduced
contamination from quasielastic events in the target windows.  The new results
are typically lower by $\sim$1\%.

The electric and magnetic form factors, $\ge(Q^2)$ and $\gm(Q^2)$, describe
the distribution of charge and magnetization in the proton. The form
factors are extracted in elastic electron--proton scattering and mapped
out as a function of the four-momentum transfer squared, $Q^2$, to yield the
momentum-space structure of the proton. Precision measurements of proton form
factors over a large kinematic range can provide important constraints on
models of the proton.  However, when extracting the form factors from
unpolarized cross section measurements using the Rosenbluth separation
technique, it is difficult to precisely separate $G_E$ from $G_M$ in the
proton for very high or very low $Q^2$ values.  The addition of polarization
measurements~\cite{akhiezer58, akhiezer68, dombey69, arnold81} allows for a
much better separation of $G_E$ and $G_M$.  Initial measurements for the
proton focused on the high-$Q^2$ region~\cite{jones00, punjabi05, gayou02, 
puckett11, puckett10}, which showed a significant falloff in the ratio
$\mugegm$ with $Q^2$, in contrast to previous extractions from Rosenbluth
separations~\cite{arrington03a}.  This difference is now believed to be due to
the contribution of two-photon exchange effects which have a large impact on
the extractions from the unpolarized cross section measurements but have less
impact on the polarization measurements~\cite{guichon03, blunden03,
arrington04d, arrington07c, carlson07, arrington11b}.  These significantly
improved measurements of $G_E$ led to a great deal of theoretical work aimed at
understanding this behavior~\cite{hydewright04, arrington07a, perdrisat07,
arrington11a}, which showed, among other things, the importance of quark
orbital angular momentum in understanding the proton structure at high
momentum~\cite{belitsky03, brodsky04a, ralston04}.  These results also had a
significant impact on studies of the correlations between the spatial
distribution of the quarks and the spin or momentum they carry, showing that
the spherically symmetric proton is formed from a rich collection of complex
overlapping structures~\cite{miller03}.

While initial investigations focused on extending proton measurements to higher
$Q^2$, the polarization measurements can also be used to improve extractions
at low $Q^2$ values, providing improved precision and less sensitivity to
two-photon exchange corrections. The low-$Q^2$ form factors relate to
large-scale structures in the proton's charge and magnetization distributions.
As such, it has long been believed that the ``pion cloud'' contributions, e.g.
the fluctuation of a proton into a virtual neutron--$\pi^+$ system, will be
important at low $Q^2$, as the mass difference means that the pion will
contribute to the large distance distribution in the bound nucleon--pion
system.  It was recently suggested that such structures are present in all the
nucleon form factors~\cite{friedrich03}, centered at $Q^2$ $\approx$ 0.3
GeV$^2$, and that these structures reflect contributions from the pion cloud
of the nucleon.  However, the significance of the proposed structures and their
interpretation as a pion cloud effect have been much disputed. This low-$Q^2$
region is also important in parity violating electron scattering
measurements~\cite{armstrong05, aniol06a, aniol06b, acha07} aimed at
investigating the strange-quark contributions to the proton electromagnetic
structure.  Isolating the strange quark contributions relies on precise
determinations of the proton form factors at low $Q^2$, including the impact
of two-photon exchange corrections~\cite{arrington07b} (discussed further
below).

While the form factors encode information on the spatial structure of
the proton, there are theoretical issues in extracting the spatial charge and
magnetization distributions, discussed in detail elsewhere~\cite{kelly02,
miller07, miller08b, miller08a, venkat11}. However, the difficulty in
extracting true rest-frame distributions for the proton does not interfere
with the comparison of form factor measurements and proton size/structure
corrections to atomic levels in hydrogen.  Extractions of the proton charge
radius~\cite{sick03, blunden05b, hill10, bernauer10, zhan11} define the proton
root-mean-square (RMS) radius as the slope of the form factor at $Q^2=0$. This
definition is consistent with the RMS radius needed in Lamb shift measurements
in hydrogen~\cite{mohr08} and muonic hydrogen~\cite{pohl10}. Corrections to
the hyperfine splitting~\cite{brodsky05, carlson08, carlson11} are also
extracted directly from the form factors. The charge radius is of particular
interest at present, due to the conflicting results between Lamb shift
measurements on muonic hydrogen~\cite{pohl10} and the electron scattering
results and measurements from the Lamb shift in electronic
hydrogen~\cite{mohr08}.

This experiment was motivated by the ideas discussed above: mapping out the
large scale proton structure, the benefit of improved precision in proton form
factors in order to extract strange quark form factors (and ultimately the
proton weak radius) from parity violating measurements, and the importance of
reducing the uncertainty in hyperfine splitting calculations arising from
proton finite-size corrections.

\section{Previous measurements}

Since the 1960s, measurements of the unpolarized cross section for elastic
e--p scattering have been used to separate $\ge$ and $\gm$. The cross
section is proportional to $(\tau\gm^2+\varepsilon\ge^2)$, where
$\tau=Q^2/4m_p^2$, and $\varepsilon=(1+2(1+(Q^2/4m_p^2))\tan^2\theta/2)^{-1}$.
Keeping $Q^2$ fixed while varying $\varepsilon$ allows for a ``Rosenbluth
separation''~\cite{rosenbluth50} of the contributions from $\ge$ and $\gm$. At
high $Q^2$, the factor of $\tau \gm^2$ dominates, as $\tau$ becomes large and
$\gm^2 \gg \ge^2$ (with $\gm/\ge=\mu_p$ at $Q^2=0$). This makes extraction of
$\ge$ difficult, as it contributes only a small, angle-dependent correction to
the larger cross section contribution from $\gm$. Similarly, in the limit of
very small $Q^2$, and thus very small $\tau$, it is difficult to isolate
$\gm$ except in the limit where $\varepsilon \to 0$, i.e. scattering angle
$\to$ 180$\deg$.

Polarization measurements are sensitive to the \textit{ratio} $\gegm$
and thus, when combined with cross section measurements, can cleanly
separate the electric and magnetic form factors, no matter how small their
contribution to the cross section becomes. It has been known
for some time~\cite{akhiezer58, akhiezer68, dombey69, arnold81} that
measurements of polarization observables would provide a powerful alternative
to Rosenbluth separation measurements, but only in the last decade or so have
the high polarization, high intensity electron beams been available, combined
with polarized nucleon targets or high efficiency nucleon recoil
polarimeters~\cite{perdrisat07,arrington07a,arrington11a}.

The first such measurements for the proton~\cite{jones00,gayou02} showed a
decrease in $\mugegm$ with $Q^2$, which differed from the existing Rosenbluth
separation measurements, which showed approximate form factor scaling, i.e.
$\mugegm \approx 1$. This discrepancy appeared to be larger than could be
explained even accounting for the scatter in the previous Rosenbluth
measurements~\cite{arrington03a}. A measurement using a modified Rosenbluth
extraction technique~\cite{qattan05} was able to extract the ratio $\mugegm$
with precision comparable to the polarization measurements, and showed a clear
discrepancy, well outside of the experimental systematics for either
technique. Experiments extending polarization measurements to higher $Q^2$
show a continued decrease of $\mugegm$ with
$Q^2$~\cite{gayou02,puckett10,puckett11}.

It was suggested that the two-photon exchange (TPE) correction may be able to
explain the discrepancy between the two techniques~\cite{guichon03,blunden03}.
While these corrections are expected to be of order $\alpha_{EM} \approx 1$\%,
they can have a similar $\varepsilon$ dependence to the contribution from
$\ge$. Because the contribution to $\ge$ is small at large $Q^2$, a TPE
correction of a few percent could still be significant in the extraction of
$\ge$. It was estimated that a TPE contribution of $\sim$5\%, with a linear
$\varepsilon$ dependence, could explain the difference~\cite{guichon03,
arrington04a}, and early calculations suggested effects of a few percent, with
just such a linear $\varepsilon$-dependence~\cite{blunden03,chen04}. These
corrections should also modify the polarized cross section measurements, but
it should be a percent-level correction in the extraction of $\gegm$, as there
is is no equivalent amplification of the effect. Including the best hadronic
calculations available yields consistency between the two techniques, and good
separation of $\ge$ and $\gm$ up to high $Q^2$~\cite{arrington07c,
arrington11b}. Comparisons of electron--proton and positron--proton scattering
can be used to isolate TPE contributions~\cite{arrington04b}, and a series of
such measurements are currently planned or underway~\cite{vepp_proposal,
e04116, kohl09}.

At low $Q^2$ values, the TPE should be well described by the hadronic
calculations~\cite{blunden05a, arrington07c}, and in fact the contributions
are small for $0.3 < Q^2 < 0.7$~GeV$^2$.  While this is a region where high
precision Rosenbluth separations are possible, measurements prior to 2010 had
relatively large uncertainties, typically 3--5\% or more on $\mugegm$.
Measurements using polarization observables in this region can provide a
significant improvement in precision, even in this low $Q^2$ regime. The
MIT-Bates BLAST experiment made measurements of $\mugegm$ using a polarized
target~\cite{crawford07} for $0.15<Q^2<0.6$~GeV$^2$, with typical uncertainties
around 2\%, about a factor of two improvement over most earlier data. The
experiment, which provided the best knowledge of the low $Q^2$ proton form
factor ratio when published, measured values below unity for
$Q^2>0.2$~GeV$^2$, but concluded that the overall results were consistent with
unity over the range of the experiment. Combined with the high-$Q^2$ JLab
data, which showed a clear deviation of $\mugegm$ for $Q^2 \gtorder
0.8$~GeV$^2$ , this suggested that the ratio was unity at very low $Q^2$ and
then began to fall somewhere in the range of 0.2--0.7~GeV$^2$.  The fact that
there was no clear indication of where the ratio began to fall below unity was
one of the motivating factors for this measurement. The updated results of this
reanalysis of \cite{ron07} provide an independent extraction of $\mugegm$ in
this kinematic region, with precision comparable to the BLAST results. More
recently, JLab experiment E08-007~\cite{zhan11}, a high-statistics follow up
to the work we present here, used the same polarization transfer techniques
but with coincident detection of the final-state electron and proton for all
kinematics, yielding measurements of $\mugegm$ with average uncertainties
below 1.2\%.

Last year, new measurements in this $Q^2$ region were also obtained by an
experiment at Mainz~\cite{bernauer10}.  The experiment made high-precision
measurements of unpolarized cross sections at $\sim$1400 kinematic points for
$Q^2<1$~GeV$^2$.  While they do not provide direct Rosenbluth extractions of
$\ge$ and $\gm$, they show a global fit to their cross section results.  Their
extraction of $\gm$ is systematically 2--4\% above previous world's data,
implying a difference of 4--8\% in the extrapolation of the cross section to
$\varepsilon=0$.  It is difficult to determine how much of their error band
could be strongly correlated in $Q^2$, as there is no information given on the
size or sources of systematic uncertainty assumed in their analysis.  While
they apply a very limited form of the two-photon exchange
corrections~\cite{mckinley48}, which is neglected in most previous
extractions, this should only reduce their value of $\gm$ relative to the
uncorrected results, implying that the true discrepancy is even larger. At
this point, it is not clear why there is such a large discrepancy between
their fit and previous measurements.

\section{Experiment details}

This experiment was carried out in Hall A of the Thomas Jefferson
National Accelerator Facility (JLab), in the summer and fall of 2006,
as part of experiment E05-103~\cite{e05103}. While the experiment was
focused on polarization observables in low energy deuteron
photodisintegration~\cite{Glister:2010ft}, elastic electron--proton scattering
measurements used to calibrate the focal plane polarimeter provided
high statistics data that allowed for an improved extraction of the proton
form factor ratio $\mugegm$ at low $Q^2$.

A polarized electron beam was incident on a cryogenic liquid hydrogen
target, nominally 10~cm in length for the 362~MeV beam energy running and 15~cm for
the 687~MeV settings (the target length was misstated as 15~cm for all runs in
the previous publication~\cite{ron07}). 
The target cells are Al, with beam entrance windows about 0.1 mm
thick, and beam exit and sides $\sim$0.2mm thick (with some variation between
the different targets).
Elastic e--p scattering events were
identified by detecting the struck proton in one of the High Resolution
Spectrometers (HRS)~\cite{alcorn04}. Data were taken with a
longitudinal polarization of approximately 40\% and with the beam helicity
flipped pseudo-randomly at 30Hz. For some settings, the scattered electron
was detected in the other HRS spectrometer.

The polarization of the struck protons is measured in a focal plane polarimeter
(FPP) in the proton spectrometer. Operation and analysis of events in the
FPP is described in detail in Refs.~\cite{alcorn04,punjabi05}. Analysis of the
angular distribution of rescattering in the polarimeter allows us to extract
the transverse polarization at the detector, which can be used to reconstruct
the longitudinal and transverse (in-plane) components of the polarization of
the elastically scattered protons. In the Born approximation, the ratio of
these polarization components is directly related to the ratio $G_E/G_M$,
\begin{equation}
R \equiv \frac{\ge}{\gm}=-
\frac{E_0+E^\prime}{2m_p}\tan\left(\frac{\theta_e}{2}\right)\frac{C_x}{C_z},
\label{eq:polxfer}
\end{equation}
where $C_{z,x}$ are the longitudinal and transverse components of the proton
polarization, $E_0$ is the beam energy, and $\theta_e$ and $E^\prime$ are the
scattered electron's angle and momentum (reconstructed from the measured
proton kinematics). Because the extraction of $\mugegm$ depends on the ratio
of two polarization components, knowledge of the absolute beam polarization
and FPP analyzing power are not necessary, although high polarization and
analyzing power improve the figure of merit of the measurement.

In the experiment, we measure the polarization not at the target, but in the
spectrometer focal plane, and the asymmetry in the rescattering is sensitive
only to polarization components perpendicular to the proton direction.  If we
look at the central proton trajectory, where the spectrometer is well
represented by a simple dipole, then the transverse component, $C_x$, will be
unchanged, while the longitudinal component, $C_z$, will be precessed in the
dipole field. If we chose a spin precession angle, $\chi$, near 90 degrees,
the longitudinal and transverse polarization components at the target will
yield ``vertical'' and ``horizontal'' components in the frame of the focal plane
polarimeter, allowing for both to be extracted by a measurement of the
azimuthal distribution of rescattering in the carbon analyzer.  In the
analysis, we use a detailed model of the spectrometer to perform the full spin
precession, rather than taking a dipole approximation, as described in detail
in Ref.~\cite{punjabi05}.

\begin{table}[ht]
\begin{center}
\caption{\label{tab:kin} Kinematics and FPP parameters for the measurements.
$\theta^{~p}_{lab}$ and $T_p$ are the proton lab angle and proton kinetic
energy, respectively. $T_{analyzer}$ is the thickness of the FPP carbon
analyzer and $\chi$ is the spin precession angle for the central trajectory.
The final column shows which kinematics had single-arm (S), coincidence (C),
or a combination of both (C/S).}
\begin{tabular}{|c|c|c|c|c|c|c|}
\hline
 $Q^2$ & $E_e$ & $\theta^{~p}_{lab}$ & $T_p$ & $T_{analyzer}$ & $\chi$ & S/C\\
(GeV$^2$) & (GeV) & (deg) & (GeV) & (inches) & (deg) & \\

\hline
0.215 & 0.362 & 28.3 & 0.120 & 0.75 & 91.0 & S \\
0.235 & 0.362 & 23.9 & 0.130 & 0.75 & 91.9 & S \\
0.251 & 0.362 & 18.8 & 0.140 & 0.75 & 92.7 & S \\
0.265 & 0.362 & 14.1 & 0.148 & 0.75 & 93.4 & S \\
0.308 & 0.687 & 47.0 & 0.170 & 2.25 & 95.3 & C \\
0.346 & 0.687 & 44.2 & 0.190 & 3.75 & 97.0 & C\&S \\
0.400 & 0.687 & 40.0 & 0.220 & 3.75 & 99.6 & S \\
0.474 & 0.687 & 34.4 & 0.260 & 3.75 & 103.0 & S \\
\hline
\end{tabular}
\end{center}
\end{table}

A follow-up experiment, JLab E08-007~\cite{e08007} was proposed to make
extremely high precision measurements in this kinematic regime. The
measurement was run in the summer of 2008, and in the analysis of the E08-007
data, it was observed that the result was somewhat sensitive to the cuts
applied to the proton kinematics when isolating elastic e--p scattering.

In this experiment, only the proton was
detected for most kinematic settings, and the elastic scattering events were
isolated using cuts on the over-determined elastic kinematics. In the original
analysis~\cite{ron07}, relatively loose cuts were applied because the
measurement was statistics limited and little cut dependence had been observed
in previous measurements~\cite{gayou01, gayou02, punjabi05}.  Most of these
measurements had high resolution reconstruction of both the proton and
electron kinematics, and so loose cuts on the combined proton and electron
kinematics provided clean isolation of the elastic peak. In addition, the
previous measurements were generally at higher $Q^2$ and so of significantly
lower statistical precision, typically 3--5\%, so it was difficult to make
precise evaluation of the impact of tight cuts on the proton kinematics.
Because the elastic events could be cleanly identified without tight cuts on
the proton kinematics, this was not considered to be a significant concern.

In the follow-up experiment, E08-007~\cite{e08007}, the electron was detected
in a large acceptance spectrometer with limited momentum and angle resolution.
The electron detection led to significant suppression of scattering from the
target windows, but the poor electron resolution required that the elastic
peak be defined using cuts on the proton kinematics. Because of this,
and the high statistics of the data set, it was possible to
make detailed studies of the cut dependence of the result. It was found that
there were small but noticeable changes in the extracted form factor ratio
if the proton kinematic cuts were made too loose, even in cases where the
endcap contributions were small.

Motivated by these issues, we reanalyzed the data from our experiment.  We
include a more careful examination of cuts used to identify the proton events
and an updated evaluation of the contribution from the target endcaps.  With
our new, more restrictive cuts, there were small but systematic changes in the
extracted form factors. These were mainly due to the reduction in the
contribution from electron scattering in the Aluminum endcaps rather than any
changes in the events corresponding to scattering from hydrogen.

One of the most important issues in the original analysis was the correction
for events that came from scattering in the Aluminum endcaps of the targets,
and there were several difficulties involved in making these corrections.
For systematic checks, we took data with the elastic peak centered on the
focal plane, but also with the spectrometer momentum approximately 2\% higher
and lower, to map out the response of the FPP across the focal plane.  Data on
the Aluminum dummy targets were typically taken for only one setting, and so
there was a systematic uncertainty associated with the stability of the size
of the background contamination and the possible variation of the polarization
transfer coefficients measured from the dummy target.  Data were taken using
both 15~cm and 10~cm cryogenic hydrogen targets, but only dummy foils for the
standard 4~cm and 15~cm cryotargets were available, yielding some additional
systematic uncertainties for the 10~cm targets.  Finally, for some runs the
beam position was not perfectly stabilized on the dummy foils and the beam,
rastered to a 4 by 4~mm$^2$ spot at the target, either partially missed the
dummy foils or impinged on both the 4~cm and 15~cm dummy targets. While the
beam position is continuously monitored and we correct for any deviation in
the event reconstruction based on the position, the luminosity is not well
known if the beam is partially missing the foils. Therefore, the relative
normalization of the contribution from the target endcaps and the dummy foils
had to be determined by looking at quasielastic events that are above the
threshold for scattering from the proton, rather than being calculated 
directly, yielding an additional systematic in the relative normalization of
the endcaps and dummy foils.

In the original analysis, endcap scattering typically yielded 3--5\% of the
cross section after all cuts were applied (much less for the two coincidence
settings), so there was a small but significant correction.  Because of the
issues mentioned above, there were very large systematic uncertainties
associated with these corrections.  In addition, the original analysis applied
the full set of cuts for elastic scattering to data from the dummy target,
yielding measurements of the polarization coefficient for endcap scattering
with extremely poor statistics and thus large fluctuations.  We now use much
tighter cuts on the reconstructed target position to try to remove most of the
endcap contributions, resulting in contributions of $\ltorder$0.5\%.  While
the cuts reduce the statistics of the main measurement somewhat, the final
uncertainty is often better, as the endcap subtraction, which had large
statistical and systematic uncertainties, is now much smaller.  We also use
looser cuts when extracting $C_x$ and $C_z$ from scattering in the aluminum
endcaps, with an extra systematic uncertainty applied to account for possible
cut dependence.

\section{Analysis details}

For elastic scattering using an electron beam with a known energy
the complete scattering kinematics can be determined from the measurement of a
single kinematic quantity, typically the angle or energy of
the final state electron or proton.  If two quantities are measured, then
the consistency of the two kinematic variables can be used to determine if
the event was associated with elastic scattering. For this
analysis, we use the proton scattering angle and momentum to reconstruct
the kinematics and to identify elastic events.  For some kinematics, the
electron was also detected, which allows for almost complete suppression of
events coming from quasielastic scattering in the aluminum entrance and
exit windows of the target.  To identify the elastic peak, we use the difference
between the measured proton momentum and the momentum calculated based on
the measured proton angle measured proton angle.  The specific variable we use
is $DpKin$, which is the momentum difference, $p_p - p_{elastic}(\theta_p)$,
divided by the central momentum setting of the proton spectrometer.  This
yields a fractional momentum deviation from the expectation for elastic
scattering.  

\begin{figure}[htb]
\begin{center}
\includegraphics[width=3.5in]{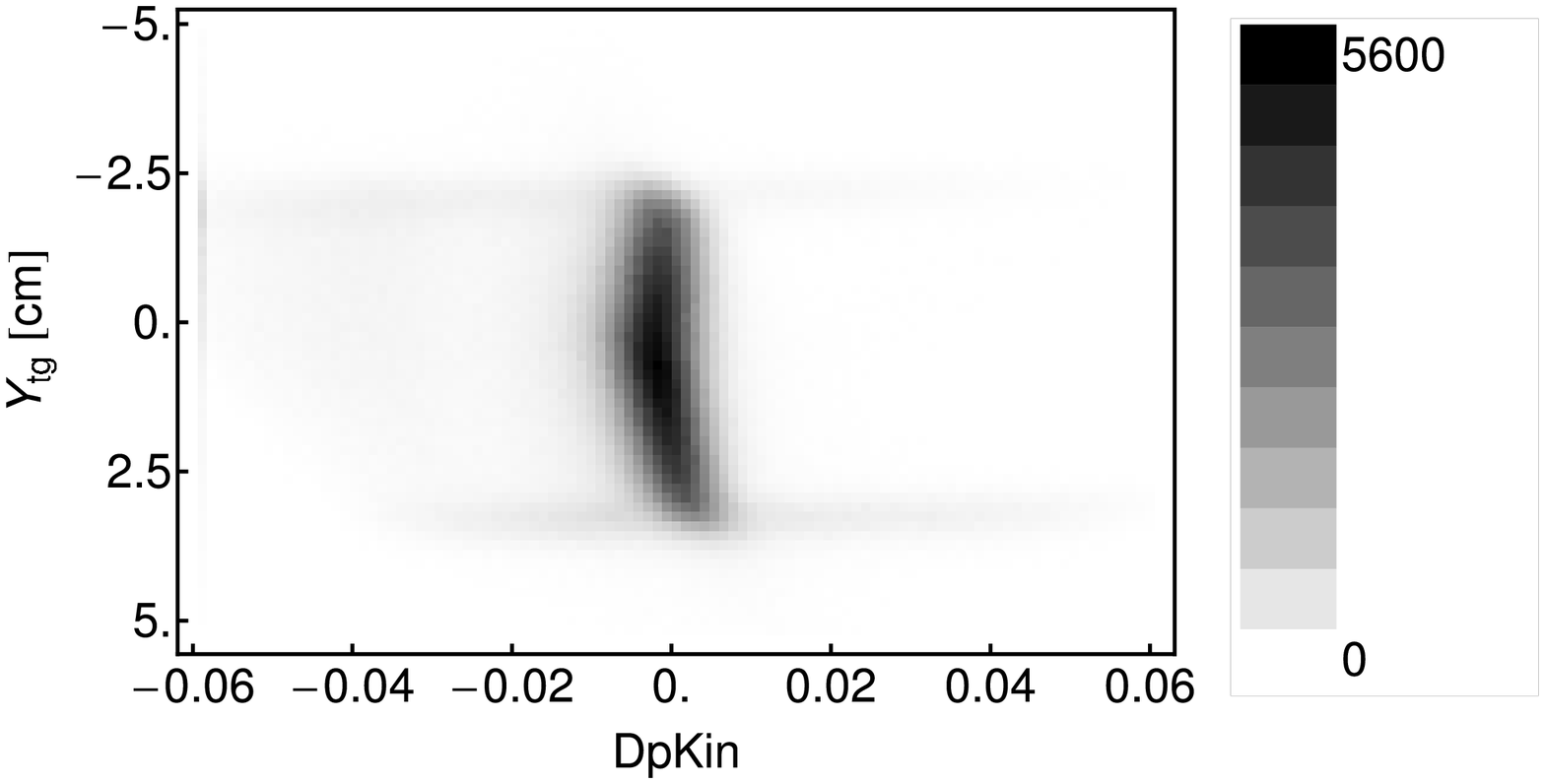}\\
\includegraphics[width=3.5in]{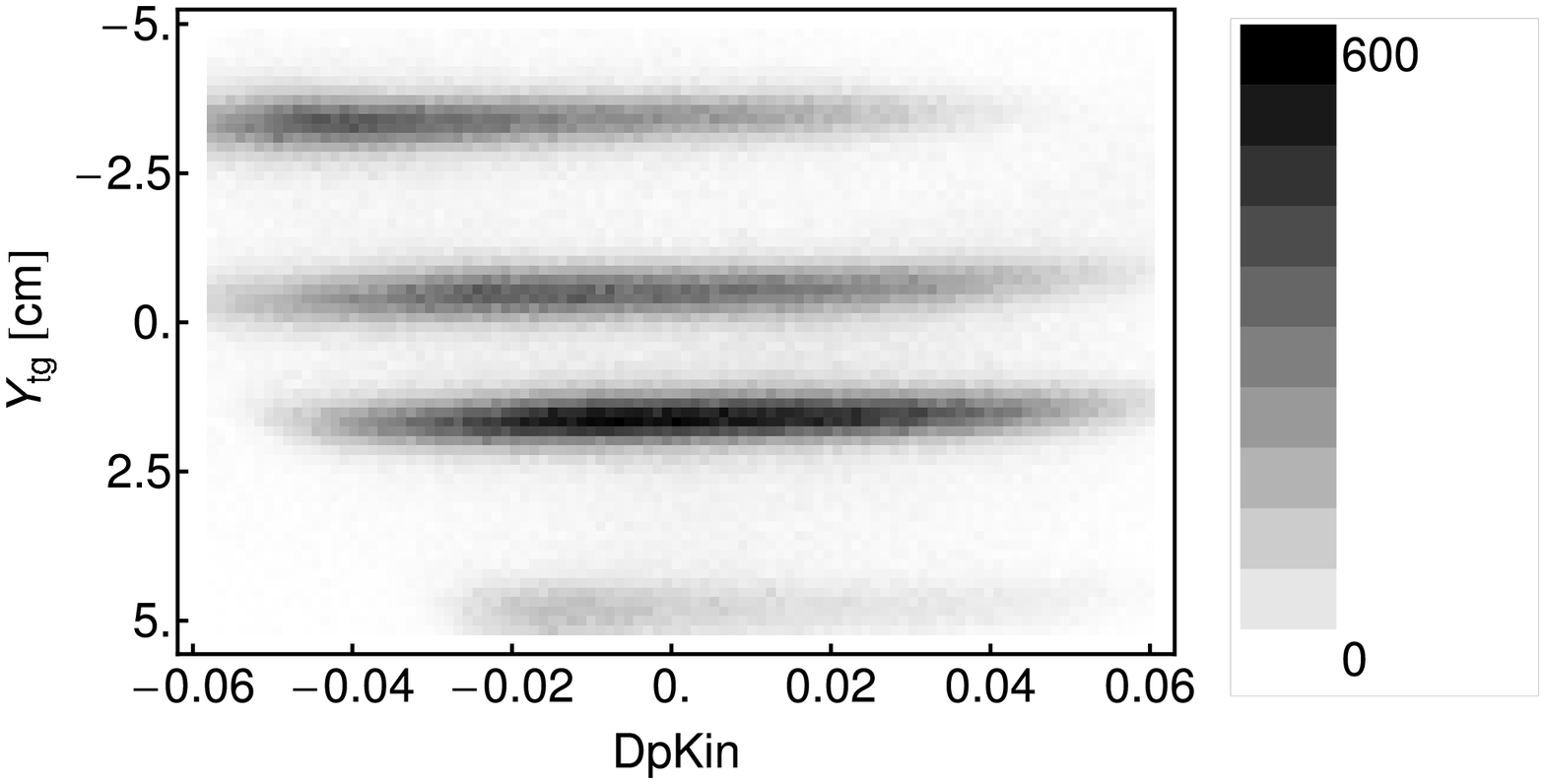}
\caption{Reconstructed target position $Y_{tg}$ vs. $DpKin =
(p_p-p_{elastic}(\theta_p))/p_{HRS}$ for the measurements on the 10~cm liquid
hydrogen target (top) and on the 4~cm and 15~cm Aluminum ``dummy'' foils
(bottom). Note that $Y_{tg}$ is the position transverse to the spectrometer
optic axis, not the position along the beamline; this difference leads to the
target dimensions being reduced by a factor of $\approx$2 here. The elastic
peak is clearly visible at $DpKin \approx 0$ for the LH2 target, while the
broad quasielastic contributions from endcap scattering are visible at the
ends of the LH2 target.}
\label{fig:targets}
\end{center}
\end{figure}

Figure~\ref{fig:targets} shows the distribution of events versus $DpKin$ and
the reconstructed target position, $Y_{tg}$, as seen by the spectrometer.  For
the hydrogen target (top panel), there is a strong peak at $DpKin \approx 0$,
corresponding to elastic events. At the extreme $Y_{tg}$ values, there is a
faint but broad distribution corresponding to quasielastic scattering in the
endcaps.  We apply a cut to $Y_{tg}$ to remove most of the contribution from
the endcap scattering, and use the measurements from the 4~cm and 15~cm dummy
target (bottom panel) to subtract the residual contribution.  Note
that for the spectra shown in Fig.~\ref{fig:targets}, the length of the LH2
target does not match either the inner or the outer pair of foils from the
dummy target. This means that the acceptance as a function of $DpKin$ depends
on $Y_{tg}$ and so will not be identical for the endcaps and the foils in the
dummy target. This is clear for the outer foils of the dummy target, where
there is a significant loss of events at extreme positive (negative) values of
$DpKin$ for the upstream (downstream) dummy foils.

\begin{figure}[htb]
\begin{center}
\includegraphics[width=3.3in,height=2.0in]{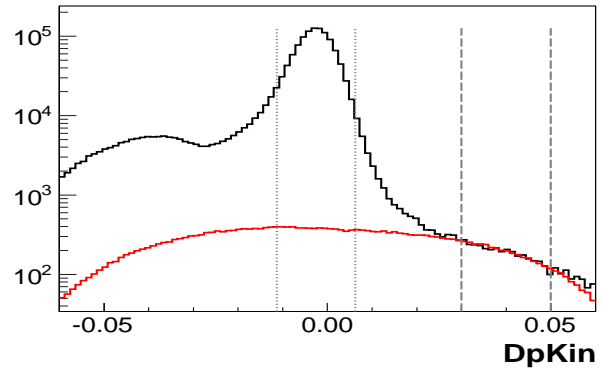}
\caption{(Color online) The $DpKin$ distribution for the hydrogen target
(thick black histogram) and the dummy targets (thin red histogram).  The
dashed vertical lines indicate the region used to normalize the dummy
contribution to match the contribution from the aluminum endcaps of the
hydrogen target and the vertical dotted lines indicate the part of the elastic
peak used in the analysis.}
\label{fig:dpkin}
\end{center}
\end{figure}

For each $Q^2$ setting, three measurements were taken; one with the elastic
peak positioned at the central momentum of the spectrometer, and two where the
elastic peak was shifted up (down) by 2\% in momentum. This allowed us to
verify that the result was independent of the position of the events on the
focal plane. However, dummy events were typically taken at only one of these
three settings, and the extracted endcap contribution and quasielastic recoil
polarizations taken from that measurement were applied to all three settings,
so the $DpKin$ distributions will not be exactly identical, especially far
away from the elastic peak. The dummy spectra are normalized to match the
observed ``super-elastic'' contribution ($DpKin > 0.03$ in
Fig.~\ref{fig:dpkin}) in the LH2 data, using only the inner foils for the
dummy target, as they have a $DpKin$ acceptance which better matches the
endcaps.  Figure~\ref{fig:dpkin} shows the spectra for the LH2 target (thick
black histogram) and the dummy target (grey histogram), after the dummy target
has been normalized in the region indicated by the vertical dashed lines.
After normalizing the spectra in this region, we can determine the endcap
contribution under the elastic peak.  The region used to define elastic events
in the analysis is indicated by the vertical dotted lines. We take a
conservative approach and apply a 50\% systematic uncertainty to the size of
the endcap contribution when making the correction for these events to account
for the impact of the different $DpKin$ spectra between the endcaps and the
dummy foils and possible variation for the settings which are shifted by
$\pm$2\% in momentum.

Having determined the contribution from endcap scattering, we use the data
from the dummy targets to determine the contributions from quasielastic
scattering to the recoil polarization components $C_x$ and $C_z$.  If we
apply the same cuts to the dummy target as we use in the analysis of the
hydrogen, there is very little data left, and we can not make a reliable
extraction of $C_x$ and $C_z$.  For the quasielastic scattering, we use all
four aluminum foils and a broader cut on $DpKin$ to determine the quasielastic
values for $C_x$ and $C_z$, and then assume that the coefficients are identical
when looking at the central part of the quasielastic spectrum.  Comparisons
showed complete consistency between the extracted values of $C_x$ and $C_z$ 
when comparing the inner and outer dummy foils for all kinematics, or when
comparing the central part of the quasielastic peak to the off-peak
contributions. Because we could not make a precise determination of $C_x$ and
$C_z$ without averaging over a larger kinematic region, we apply an
uncertainty to $C_x$ and $C_z$ of 0.02 and 0.05 respectively, compared to
typical values for these polarization components in this experiment of
0.08--0.2 for $C_x$ and 0.15--0.3 for $C_z$.

In the original analysis~\cite{ron07}, the $Y_{tg}$ cut was loose and so there
was a large (3--5\%) contribution from endcap scattering which had to be
subtracted.  Because the tight cuts used on the elastic events were also
applied to the dummy spectra used to subtract endcap scattering contributions,
the statistical uncertainty on these subtractions could be very large. 
Therefore, fluctuations in the low statistics dummy measurements led to large
uncertainties and significant fluctuations in the dummy-subtracted
measurements.  In the present analysis, the endcap contributions are greatly
reduced, with a maximum contribution well below 1\%, such that the
conservative systematic uncertainties assumed for the dummy normalization and
polarization coefficients yield only small uncertainties in the final result. 
While the tighter cuts yield slightly reduced statistics in the elastic peak,
the total statistical uncertainty is sometimes smaller because the 
background contribution was reduced. Note that for a few settings, additional
runs were included, improving the statistics by 5--15\%, but this was a small
effect compared to the modified cuts.

There were also some small changes in the evaluation of the systematic
uncertainties. In the previous analysis, the systematic uncertainty from
the endcap contribution was folded into the reported statistical
uncertainties, and these are now part of the quoted systematics.  In addition,
the estimated systematics are somewhat larger than in the previous analysis,
due to a more detailed analysis of the uncertainty in the spin precession
through the spectrometer~\cite{xiaohui_thesis}.

\begin{figure}[htb]
\begin{center}
\includegraphics[height=2.0in,width=3.3in]{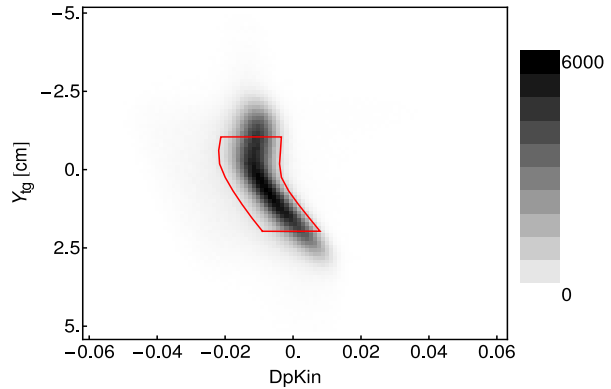}\\
\caption{(Color online) The reconstructed target position, $Y_{tg}$ vs.
$DpKin$, the deviation of the momentum from the expected elastic peak
position.  A correction for the average energy loss is applied, but there is a
significant difference for events on the upstream side of the target, which
exit through the side wall of the target, and events which occur nearer the
downstream end of the target and have less energy loss.  The band indicates
the graphical cut placed on these runs, to approximate a two-sigma range for
each $Y_{tg}$ value.}
\label{fig:2dcut}
\end{center}
\end{figure}

The proton energy loss, which can be significant for the low $Q^2$ kinematics,
was also more carefully evaluated, leading to a small change in the average
$Q^2$ for each bin. For the 362~MeV running, where the proton was detected at
small angles, the energy loss depends on the position in the target where the
scattering occurs. Figure~\ref{fig:2dcut} shows $DpKin$ vs. $Y_{tg}$ for one
of the 362~MeV runs with an average proton energy loss is applied to all
events. Positive $Y_{tg}$ values correspond to the upstream portion of the
target, where all events exit through the side of the target and travel
through a constant amount of hydrogen and aluminum and can be well corrected
assuming a fixed energy loss. Events that exit through the downstream end of
the target lose less energy because they pass through less material, yielding a
$Y_{tg}$-dependent position for the elastic peak.  This yields a reduced
proton energy loss, and thus a higher apparent proton momentum, for events
that occur near the exit window.  For the kinematics where this effect is
important, we apply a $Y_{tg}$-dependent cut, cut corresponding to a two-sigma
region around the elastic peak for each region of $Y_{tg}$, as indicated by
the graphical cut displayed in Fig.~\ref{fig:2dcut}.  The reconstructed value
of $DpKin$ is only used to select elastic events, so while a position-dependent
energy loss could have been applied, one would still end up with the same set
of good events passing the cuts.  In our approach, we are not sensitive to any
imperfections in the energy loss correction, since we use a two-sigma cut for
all $Y_{tg}$ values.

\begin{table}[ht]
\begin{center}
\caption{\label{tab:cuts} Kinematic-dependent cuts applied to the data.
The $Y_{tg}$ cut is chosen to significantly suppress any contributions
from the target endcap (as shown in Fig.~\ref{fig:2dcut}).}
\begin{tabular}{|c|c|c|c|}
\hline
 $Q^2$    & $\theta^{~p}_{lab}$ & $Y_{tg}$ cut & $\delta p/p$ cut \\
(GeV$^2$) & (deg)               &   (cm)       &                  \\
\hline
0.215~&~28.3~&~$-0.022<Y_{tg}<0.018$~&~$|\delta p/p|<0.045$~\\
0.235~&~23.9~&~$-0.022<Y_{tg}<0.018$~&~$|\delta p/p|<0.045$ \\
0.251~&~18.8~&~$-0.018<Y_{tg}<0.012$~&~$|\delta p/p|<0.045$ \\
0.265~&~14.1~&~$-0.014<Y_{tg}<0.010$~&~$|\delta p/p|<0.045$ \\
0.308~&~47.0~&~$-0.025<Y_{tg}<0.020$~&~$|\delta p/p|<0.040$ \\
0.346~&~44.2~&~$-0.025<Y_{tg}<0.020$~&~$|\delta p/p|<0.040$ \\
0.400~&~40.0~&~$-0.028<Y_{tg}<0.022$~&~$|\delta p/p|<0.040$ \\
0.474~&~34.4~&~$-0.024<Y_{tg}<0.020$~&~$|\delta p/p|<0.040$ \\
\hline
\end{tabular}
\end{center}
\end{table}

The kinematic-dependent cuts are detailed in Table~\ref{tab:cuts}.  In
addition, several cuts were applied to all kinematics.  A cut was applied
on the out-of-plane angle, $|\theta_{tg}|<0.06$~rad, and the in-plane angle,
$|\phi_{tg}|<0.03$~rad, to ensure events were inside of the angular acceptance
of the spectrometer.  A two-sigma cut was applied on the $DpKin$ peak, with
a $Y_{tg}$-dependence cut for the low energy kinematics to account for the
position-dependent average energy loss as shown in Figure~\ref{fig:2dcut}.
The tracks before and after the Carbon analyzer were used to determine the
scattering location and scattering angle in the analyzer.  Events were required
to have the secondary scattering occur within the analyzer, and angle between
5 and 50$\deg$ were accepted.  In addition, we apply a cone test~\cite{punjabi05} to
ensure that there is complete azimuthal acceptance in the FPP.  We do this
by requiring that the FPP would have accepted events with any azimuthal angle
given the reconstructed vertex and scattering angle.  This ensures that 
any asymmetry in the acceptance or distribution of events does not lead
to a difference in the scattering angle distribution for vertical and horizontal
rescattering.  A significant difference between the rescattering distribution
for vertical and horizontal rescattering events would yield a different average
analyzing power, and the analyzing power would not cancel out in the ratio
of polarization components.

\begin{table}[tb]
\begin{center}
\caption{\label{tab:sys} Systematic uncertainties on $R=\mugegm$. See text
for details.}
\begin{tabular}{|c|c|c|c|}
\hline
 $Q^2$	 & $\delta R$	& $\delta R$	& $\delta R$	\\
(GeV$^2$)& (endcap)	& (optics) 	&  (cuts)	\\
\hline
~ 0.215	~&~ 0.0012	& 0.0079	& 0.0141	\\
~ 0.235	~&~ 0.0004	& 0.0079	& 0.0120	\\
~ 0.251	~&~ 0.0003	& 0.0078	& 0.0107	\\
~ 0.265	~&~ 0.0003	& 0.0076	& 0.0098	\\
~ 0.308	~&~ --		& 0.0091	& 0.0077	\\
~ 0.346	~&~ --   	& 0.0086	& 0.0066	\\
~ 0.400	~&~ 0.0010	& 0.0088	& 0.0056	\\
~ 0.474	~&~ 0.0007	& 0.0117	& 0.0049	\\
\hline
\end{tabular}
\end{center}
\end{table}

The combination of the more restrictive cuts on the elastic events and the
associated reduction in contamination due to scattering from the target
windows leads to a reduction in the extracted ratio that is typically at the
1--2\% level. The largest effect is due to the improved correction for endcap
scattering, mainly due to cuts which significantly reduced the size of this
contribution. There is also a ~1\% reduction in the coincidence settings,
where there are negligible endcap contributions, which is due to the tighter
cuts on the proton kinematics. Tight elastic kinematics cuts using just
the proton will remove events where there is a larger than average
error in the reconstruction of the proton scattering angle or momentum due to
multiple scattering or imperfect track reconstruction. While these errors are
small, the reconstructed kinematics are used to determine the spin propagation
through the spectrometer, and thus the impact of the poor reconstruction may
be amplified in evaluating the spin precession.  Table~\ref{tab:sys} shows
the various contributions to the systematic uncertainty as a function of
$Q^2$.  At high $Q^2$, the uncertainty in the spin precession due to 
imperfect knowledge of the spectrometer optics dominates.  At low $Q^2$,
the uncertainty is dominated by our ability to determine the cut dependence of
the result.  The cut-dependent uncertainties come mainly from two sources;
possible variation of the result due to the cuts on $y_{tg}$ and $DpKin$.
While no systematic cut dependence with the $y_{tg}$ cut was observed, we
apply a 0.4\% uncertainty as a conservative estimate based on examining the
variation of $\mugegm$ with the $y_{tg}$ cut, in particular for the
coincidence data where the background contributions are smaller.  For $DpKin$,
we estimate the uncertainty based on varying the width of the cut around the
elastic peak.  This was done for both these data and the E08-007 results and
no systematic cut dependence was observed, so the scatter of the results was
taken as a conservative estimate of the systematic uncertainties. Because the
data taken at low beam energy do not have sufficient statistics to set precise
limits, we fit the uncertainties from the higher energy measurements and the
E08-007 results and find a behavior consistent with $1/Q^4$, which we use to
obtain the quoted uncertainties for the low $Q^2$ values.

\section{Results}

\begin{figure}[htb]
\begin{center}
\includegraphics[width=3.3in]{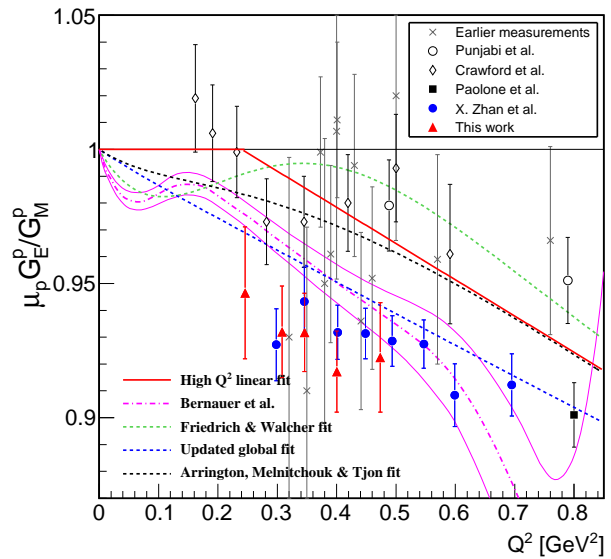}
\caption{(Color online) The proton form factor ratio as a function of
$Q^2$ (with the four low-$Q^2$ measurements combined into one data point)
shown with previous extractions with total uncertainties $\leq$3\%. The curves
are various fits~\cite{gayou02, friedrich03, zhan11, arrington07c}, while the
dot-dashed curve and associated error bands show the result of the fit to the
recent Mainz measurements~\cite{bernauer10}.}
\label{fig:data1} 
\end{center}
\end{figure}

The results of the reanalysis are given in Table~\ref{tab:res} and shown in
Figure~\ref{fig:data1}, which presents the updated results along with previous
measurements and a selection of fits.  The updated analysis yielded a
systematic decrease of $\sim$1\% in the extracted ratio, except for the
highest $Q^2$ point which decreased by 5\%. The analyzing power has been
extracted from these data~\cite{glister09}, but the quality of this
extraction does not impact these results, as the analyzing power cancels out
in the ratio of Eq.~\ref{eq:polxfer}.  However, because the FPP efficiency and
analyzing power are significantly lower for the data taken at a beam
energy of 362~MeV (due to the lower proton momentum and thinner analyzer),
the statistical uncertainty in these points is much larger. The two-photon
exchange corrections from the hadronic calculation of Blunden, et
al.~\cite{blunden05a} are 0.35\% for the data below $Q^2=0.3$~GeV$^2$ and
0.2\% for the higher $Q^2$ points.  This is well below the statistical and
systematic uncertainties for all points, and no correction (or uncertainty)
for the TPE effects is included in the extraction.

\begin{table}[tb]
\begin{center}
\caption{\label{tab:res} Experimental Results. $R$ is given along with its
statistical and systematic uncertainties. The last column (f) is the fractional
contribution from scattering in the target endcaps, along with the statistical
uncertainty; a 50\% systematic uncertainty is also applied.  The contribution is
negligible for the coincidence settings.  For $Q^2=0.474$~GeV$^2$, dummy
measurements were taken at all three sub-settings, and the range of results is
given. $^\dag$The final entry is the average of the four low-statistics
point below $Q^2=0.3$~GeV$^2$.}
\begin{tabular}{|c|c|c|}
\hline
 $Q^2$	 & $R=\mugegm$	& f	 \\
(GeV$^2$)&		& (\%)	 \\
\hline
~ 0.215	~&~	0.8250~$\pm$~0.0483~$\pm$~0.0162 ~& 0.26(3)   \\
~ 0.235	~&~	0.9433~$\pm$~0.0414~$\pm$~0.0144 ~& 0.13(2)   \\
~ 0.251	~&~	0.9882~$\pm$~0.0420~$\pm$~0.0132 ~& 0.19(3)   \\
~ 0.265	~&~	0.9833~$\pm$~0.0349~$\pm$~0.0124 ~& 0.16(2)   \\
~ 0.308	~&~	0.9320~$\pm$~0.0123~$\pm$~0.0119 ~& --	      \\
~ 0.346	~&~	0.9318~$\pm$~0.0098~$\pm$~0.0108 ~& --/0.40(2)\\
~ 0.400	~&~	0.9172~$\pm$~0.0109~$\pm$~0.0105 ~& 0.65(4)   \\
~ 0.474	~&~	0.9225~$\pm$~0.0160~$\pm$~0.0127 ~& 0.4--0.6  \\
\hline
\hline
~ 0.246$^\dag$	~&~	0.9465~$\pm$~0.0204~$\pm$~0.0137 ~& n/a		\\
\hline
\end{tabular}
\end{center}
\end{table}

The results in Fig.~\ref{fig:data1} show that the original conclusions
of~\cite{ron07} are largely unaffected. The new results support even more
clearly the conclusion that the ratio $\mugegm$ is below one even for these low
$Q^2$ values, with the change from previous Rosenbluth separations being
driven mainly by a change in $\ge$, with a smaller change in $\gm$. The
previous hint of a local minimum near $Q^2=0.35-0.4$~GeV$^2$ was a consequence
of the point near 0.5~GeV$^2$, and there is no longer any indication for this
in our measurement. These results further support the observation that the
decrease of the ratio below unity occurs at low $Q^2$, and thus we expect that
there will be a slightly larger impact on the extraction of strange quark
contributions, as discussed in the original paper~\cite{ron07}.

A comparison of the high-precision measurements at low-$Q^2$ shows some
small but systematic differences.  The results from the Mainz cross section
measurements~\cite{bernauer10} are 1--2\% above the recoil polarization
measurements from this work and the lower $Q^2$ results from the
recent JLab E08-007 measurement~\cite{zhan11}, although they are in agreement
with the E08-007 results at higher $Q^2$ values.  One concern for the
results extracted from the Mainz cross section measurements is the sensitivity
to two-photon exchange (TPE) corrections~\cite{arrington11b}. For the
kinematics of the Mainz experiment, these corrections are fairly small,
$\ltorder$2\%, but this is very large compared to the statistical
($\ltorder$0.2\%) and systematic ($\ltorder$0.5\%) uncertainties applied in
the global fit to $\ge$ and $\gm$. Thus, if ignored, this could yield
significant corrections compared to the quoted uncertainties. Coulomb
corrections were applied using the prescription of McKinley and
Feshbach~\cite{mckinley48}, which corresponds to the $Q^2=0$ limit of the
Coulomb distortion correction (the soft-photon approximation of the full TPE
corrections). However, over much of the $Q^2$ range of the experiment,
applying the $Q^2=0$ correction is worse than neglecting the correction
altogether, as the Coulomb correction changes sign at $Q^2\approx
0.15$~GeV$^2$~\cite{arrington04c}. An estimate of the impact of these
corrections~\cite{arrington11c} suggests that an improved prescription would
decrease $\mugegm$ extracted from the cross sections by 1--3\% for $Q^2
\gtorder 0.1$~GeV$^2$ in a direct Rosenbluth separation, although a more
complete analysis is required to determine the impact on their global fit. 
Bernauer~\etal, have examined the impact of these
corrections~\cite{bernauer11} in more detail, although the TPE prescription
they apply~\cite{borisyuk07} is only valid up to $Q^2=0.1$~GeV$^2$.  They find
that their extracted value for $\mugegm$ changes by more than the total quoted
uncertainty for all $Q^2$ values up to 0.1~GeV$^2$, and while they find
smaller changes at larger $Q^2$ values, this is where the prescription is not
expected to be valid.

If the Coulomb corrections bring the Rosenbluth~\cite{bernauer10} extractions
into agreement with the recoil polarization data, there is still a small
systematic disagreement between these and the polarized target measurements
from BLAST~\cite{crawford07}. At this point, we are unaware of any theoretical
argument that would explain a difference between the results of the two
different polarization techniques. This discrepancy can be further examined in
the second phase of the JLab E08-007 experiment~\cite{e08007} which will make
extremely high precision measurements of $\mugegm$ down to $Q^2 \approx
0.015$~GeV$^2$, allowing for a comparison with the BLAST measurements using
the same basic technique.

\begin{figure}[htb]
\begin{center}
\includegraphics[width=3.3in]{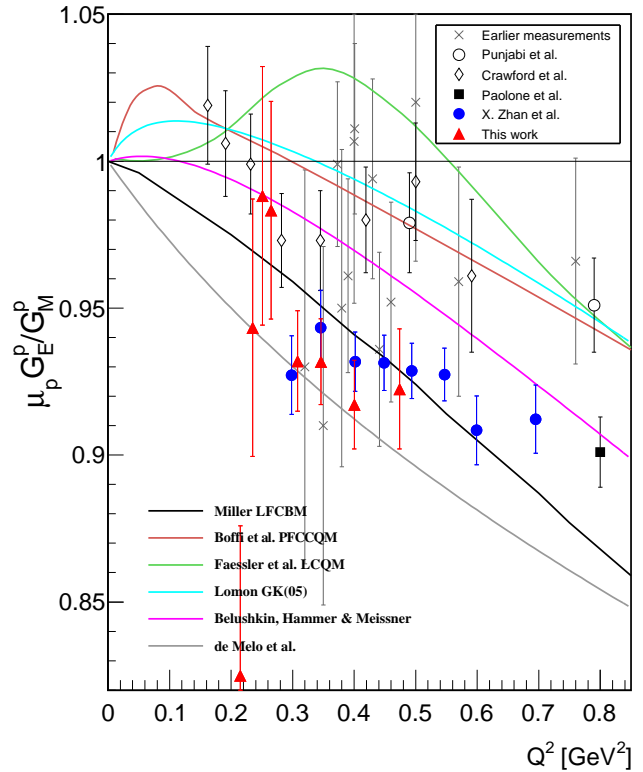}
\caption{(Color online) The proton form factor ratio $\mugegm$ vs.
$Q^2$, compared to several low-$Q^2$ models.  The curves shown are Miller's
light-front cloudy-bag model calculation~\cite{miller02}; Boffi's point-form
chiral constituent quark model calculation~\cite{boffi02}; Faessler's
light-front quark model calculation~\cite{faessler06}; Lomon's vector-meson
dominance model~\cite{lomon06}; the dispersion analysis of Belushkin, Hammer and
Meissner~\cite{belushkin07}; and the model of de Melo~\etal~\cite{demelo09}.}
\label{fig:data2} \end{center}
\end{figure}

Figure~\ref{fig:data2} shows the measurements compared to a set of theoretical
curves.  The first type of calculation is based on the constituent quark
models, which was quite successful in describing the ground state baryon
static properties. To calculate the form factors, relativistic effects need to
be considered. Miller~\cite{miller02} performed a calculation in the
light-front dynamics including the effect from the pion cloud. Boffi {\it et
al.}~\cite{boffi02} performed a point form calculation in the
Goldstone boson exchange model with point-like constituent quarks.
Faessler~\etal~\cite{faessler06} used a chiral quark model where pions are
included perturbatively and dress the bare constituent quarks by mesons in a
Lorentz covariant fashion. Another group of calculations is based on the
Vector Meson Dominance (VMD) picture, in which the scattering amplitude is
written as an intrinsic form factor of a bare nucleon multiplied by an
amplitude derived from the interaction between the virtual photon and a vector
meson. This type of models usually involve a number of free parameters for the
meson mass and coupling strength. Lomon~\cite{lomon01,lomon06,crawford10}
performed the VMD fits by including additional vector mesons and pQCD
constraints at large $Q^2$. Belushkin~\etal~\cite{belushkin07} performed a
calculation using dispersion relation analysis with additional contribution
from $\rho\pi$ and $K\bar{K}$ continua. More recently, de Melo {\it et
al.}~\cite{demelo09} performed a calculation in the light-front VMD model by
considering the non-valence contribution of the nucleon state. While most of
the theoretical curves are a few percent higher, the calculations of
Miller~\cite{miller02} and de Melo~\etal~\cite{demelo09} generally reproduce
the large deviation from $\mugegm=1$ in this low $Q^2$ region, emphasizing the
pion cloud or non-valence effect.

Our new results, as well as other high-precision measurements at low $Q^2$,
show that $\mugegm<1$ even down to very low values of $Q^2$.  A global
fit~\cite{zhan11} to the cross section and polarization measurements in this
region, including the data presented in this work, indicates that $\ge$ is
$\sim$2\% below previous fits that did not include the low $Q^2$ polarization
measurements, while $\gm$ is approximately 1\% higher. These small changes in
the low $Q^2$ form factors impact other measurements as well.  For example, it
was recently pointed out~\cite{higinbotham10} that the reduction in the form
factor yields an agreement between studies of the asymmetry in the
D(e,e$^\prime$p)n reaction at low missing momentum in polarized target
measurements at NIKHEF and MIT-Bates~\cite{passchier02, milbrath98,
milbrath99}, and recent measurements at Jefferson Lab~\cite{Hu06}.  Similarly,
this small shift in $G_E$ and $G_M$ at low $Q^2$ modifies the expected
asymmetry in parity-violating elastic electron--proton scattering, which
serves as the baseline when extracting the strange-quark contribution to the
proton form factors~\cite{armstrong05, aniol06a, aniol06b, acha07}.  The
effect is relatively small for any given extraction, especially at forward
angles where there is a partial cancellation due to the changes in $G_E$ and
$G_M$. However, because this is a systematic correction to all such
measurements, the updated form factors could have a small net contribution on
the extracted strange-quark contributions.

The form factors at very low $Q^2$ are important in both extracting the proton
size and as input to other finite-size corrections in atomic physics, e.g. the
Zemach radius in hyperfine splitting~\cite{brodsky05, carlson08, carlson11}.
Measurements of the ratio $\mugegm$ are not, by themselves, sufficient
to extract the proton radius or calculate the finite-size corrections, as 
these require the individual form factors $\ge(Q^2)$ and $\gm(Q^2)$.  However,
the polarization measurements can be used to improve global fits to cross
section data, in which determining the relative normalization of different
data sets is often the limiting factor in the systematic
uncertainties~\cite{sick03}. It is often possible to modify the normalization
of certain data sets and modify the ratio $\mugegm$ in such a way that the
cross section measurements are still relatively well fit.  Thus, having direct
constraints on $\mugegm$ with high precision allows for a better determination
of these normalization factors, and an improved extraction of the form
factors.  The fit of Ref.~\cite{arrington07c} was updated in
Ref.~\cite{zhan11} to include the data presented here as well as additional
polarization measurements~\cite{zhan11, paolone10, puckett10}, yielding
improved extractions of $\ge$, $\gm$, as well as the proton charge and
magnetization radii.  While these polarization measurements do not go as low
in $Q^2$ as one would like for the extraction of the radius, they nonetheless
play an important role in the extraction, mainly by providing improved
determinations of the relative normalization of the different experiments.  In
addition, higher order terms in the $Q^2$ expansion need to be
constrained to obtain a measure of the radius~\cite{sick03}.

This updated global analysis yielded an RMS charge radius of 0.875(10)~fm,
consistent with the CODATA value~\cite{mohr08} and recent extraction from
cross section measurements from Mainz~\cite{bernauer10}. Combined, these
independent extractions based on the electron--proton interaction yield a
radius of 0.8772(46)~fm~\cite{zhan11}, more than seven standard deviations
from the recently published PSI muonic hydrogen Lamb shift
measurement~\cite{pohl10} of 0.8418(7)~fm.  While higher order corrections,
which depend on the values of the form factors at finite $Q^2$, can modify the
extracted radius, these corrections appear to be far too small to explain the
discrepancy, although this is still under investigation~\cite{derujula10,
cloet11, distler11, wu11}.  A recent work~\cite{miller11} has proposed a
possible mechanism to explain a difference between electronic and muonic
probes of the proton structure, due to off-shell effects in the hadronic
intermediate state in the two-photon exchange diagrams.  However, while this
is an area that has received a great deal of attention in the recent
past~\cite{barger10, jentschura11, derujula10, cloet11, derujula11b,
distler11, hill11, carroll11}, the question is still unresolved.

The updated global fit of the low $Q^2$ data yields a magnetic RMS radius of
0.867(20)~fm~\cite{zhan11} .  This is in reasonable agreement with other
extractions, 0.85(3)~fm from Kelly~\cite{kelly02} and 0.857 from Hammer and
Meissner~\cite{hammer04b}. However, neither of these included Coulomb
distortion corrections, and they are global fits that include a large body of
high-$Q^2$ data which do not contain useful information on the proton radius,
and thus may not be reliable when it comes to examining the radius.
Sick~\cite{sick03} extracted the Zemach radius, which depends on both the
charge and magnetic distributions, and converting this to a magnetic radius
yields a value of 0.855(35)~fm~\cite{hammer04b}, in agreement with the updated
fit.  However, the data used by Sick are also included in this global fit, so
these cannot be considered independent extractions.  The Mainz
result~\cite{bernauer10} is 0.777(17)~fm, well below the other electron
scattering analyses, but the strong $Q^2$ dependence of the Coulomb distortion
at very low $Q^2$, neglected in their extraction, may have a significant
impact on the extraction of the radius~\cite{arrington11c}.  An updated
estimate yields a 1.5$\sigma$ shift in their magnetic radius (and almost
no change in the electric radius), yielding 0.803(17)~fm~\cite{bernauer11},
where the uncertainty does not include any contribution related to the
two-photon exchange corrections.  This increased radius is now only 2.4$\sigma$
from our result, but this updated value still appears to be an
underestimate, as their weighted averaging of results from different
fits~\cite{bernauerphd} leads to a bias towards fits with fewer parameters.
They give systematically lower values for the magnetic radius than the fits
with more parameters, as one might expect given that the precision of the
extraction of $\gm$ tends to improve as $Q^2$ increases, and so inclusion of
the high-$Q^2$ data can dominate the fit if there is not enough flexibility in
the fit.  This does not appear to affect their extraction of the charge
radius, as the extracted radius does not show the same systematic dependence
on the number of fit parameters.  This is not surprising, as precise
extractions of $\ge$ are not limited to higher $Q^2$ values.

Finally, one can use the discrepancy between calculations and measurements of
the hyperfine splitting in the hydrogen ground state to extract the magnetic
radius, assuming that all of the other proton structure corrections, including
the charge radius, are well known. The analysis by
Volotka~\etal~\cite{volotka05} yields a radius of 0.778(29)~fm, in agreement
with the Mainz result~\cite{bernauer10}.  However, an updated
calculation~\cite{carlson08} which uses updated proton form
factors~\cite{arrington07c} and spin structure is also consistent with the
hyperfine splitting even though it takes a larger magnetic RMS radius of
0.858~fm, consistent with our results.  Thus, it appears that the
hyperfine splitting does not have the sensitivity to precisely constrain
the magnetic radius, given the current uncertainty in the other portions
of the proton structure corrections.

So while the charge radius has a clear discrepancy between muonic hydrogen
lamb shift measurements and the various measurements using the
electron--proton interaction, the situation for the magnetic radius is less
clear.  The extractions of the magnetic radius are more sensitive to small
corrections, both for electron scattering and extractions from hyperfine
splitting measurements, making it difficult to tell at this point if there
are, in fact, real discrepancies between the techniques.

\section{Summary and Conclusions}

In summary, we present an updated extraction of the form factor ratio
$\mugegm$ from the data of Ref.~\cite{ron07}. We find a somewhat lower value
for $\mu_P G_E/G_M$ than the initial extraction for the entire dataset,
consistent with two recent high precision
measurements~\cite{bernauer10,zhan11}.  The new analysis does not change our
previous conclusion, i.e., that there is clear indication of a ratio smaller
than unity, even for low $Q^2$, indicating the necessity of including
relativistic effects in any calculation of the form factors in this region. 
Both the form factors and proton charge radius extraction from various e--p
scattering measurements are in agreement, and there is still a significant
disagreement with the charge radius as extracted from muonic
hydrogen~\cite{pohl10}.

\begin{acknowledgments}

This work was supported by the U.S. Department of Energy, including contract
DE-AC02-06CH11357, the U.S. National Science Foundation, the Israel Science
Foundation, the Korea Research Foundation, the US-Israeli Bi-National
Scientific Foundation, and the Adams Fellowship Program of the Israel Academy
of Sciences and Humanities. Jefferson Science Associates operates the Thomas
Jefferson National Accelerator Facility under DOE contract DE-AC05-06OR23177.

\end{acknowledgments}

\bibliography{ledex_update}

\end{document}